\documentstyle[12pt]{article}

\def\beq{\begin{equation}}
\def\eeq{\end{equation}}
\def\beqa{\begin{eqnarray*}}
\def\eeqa{\end{eqnarray*}}

\def\za{\alpha}
\def\zb{\beta}

\begin{document}

\begin{flushright}
UR-1578\\
ER/40685/937\\
Jul 1999
\end{flushright}

\vspace*{.2in}

\begin{center}
{\bf  Neutrino Oscillations from 
Supersymmetry without R-parity 
--- Its Implications on the Flavor Structure of the Theory}$^\star$

\vspace*{.4in}
{\bf Otto C.W. Kong }

\vspace*{.4in}
{\it Department of Physics and Astronomy,\\
University of Rochester, Rochester NY 14627-0171.}

\vspace*{.8in}
{Abstract}\\
\end{center}
We discuss here  some flavor structure
aspects of the complete theory of supersymmetry without R-parity addressed 
from the perspective of fitting neutrino oscillation data based on the 
recent Super-Kamiokande result. The single-VEV parametrization
of supersymmetry without R-parity is first reviewed, illustrating some
important features not generally appreciated. For the flavor structure
discussions, a naive,  flavor model independent, analysis is presented,
from which a few interesting things can be learned.

\noindent

\vfill
\noindent --------------- \\
$^\star$ Invited talk at NANP 99 conference, 
  Dubna (Jun 28 - Jul 3)
 --- submission for the proceedings.  
 
\clearpage
\addtocounter{page}{-1}

\begin{center}
{\bf  Neutrino Oscillations from 
Supersymmetry without R-parity 
--- Its Implications on the Flavor Structure of the Theory}

\vspace*{.2in}
{\bf Otto C.W. Kong }

\vspace*{.2in}
{\it Department of Physics and Astronomy,\\
University of Rochester, Rochester NY 14627-0171.}

\vspace*{.3in}
{Abstract}\\
\end{center}
\noindent
We discuss here  some flavor structure
aspects of the complete theory of supersymmetry without R-parity addressed 
from the perspective of fitting neutrino oscillation data based on the 
recent Super-Kamiokande result. The single-VEV parametrization
of supersymmetry without R-parity is first reviewed, illustrating some
important features not generally appreciated. For the flavor structure
discussions, a naive,  flavor model independent, analysis is presented,
from which a few interesting things can be learned. 

\section{Introduction and outline.}
We discuss here a simple and specific issue ---  some flavor structure
aspects of the complete theory of supersymmetry without R-parity addressed 
from the perspective of fitting neutrino oscillation data.
We will first review our formulation of supersymmetry without R-parity
and its application to study of neutrino masses. The formulation has
been reported in Ref.\cite{I}. It is based on a specific choice of flavor 
bases that allows the maximal simplification of the tree level fermion mass
matrices, as well as a comprehensive treatment of all the R-parity violating (RPV)
couplings together without any assumption. We will go on then to discuss
a simple scenario of three neutrino masses and mixings inspired by the
recent Super-Kamiokande (Super-K) result\cite{SuperK}, incorporating it into our framework of
supersymmetry without R-parity. Our concentration here is at its
implication on the flavor structure of the theory. We will discuss a
naive,  flavor model independent, analysis from which a few interesting
things can be learned. The discussion is mainly based on results presented
in Ref.\cite{K}.

\section{ Obtaining the supersymmetrized standard model.}
Let us start from the beginning and look carefully at the supersymmetrization of the 
standard model. In the matter field sector, all fermions and scalars have to
be promoted to chiral superfields containing both parts. It is 
straightforward for the quark doublets and singlets, and also for the leptonic
singlet. The leptonic doublets, however, has the same quantum number as
the Higgs doublet that couples to the down-sector quarks. Nevertheless, one
cannot simply get the Higgs, $H_d$, from the scalar partners of the leptonic 
doublets, $L$'s. Holomorphicity of the superpotential requires a separate superfield to 
contribute the Higgs coupling to the up-sector quarks. This $\hat{H}_u$ 
superfield then contributes a fermionic doublet, the Higgsino, with non-trivial
gauge anomaly. To cancel the latter, an extra fermionic doublet with
the quantum number of $H_d$ or $L$ is needed. So, the result is that we need
four superfields with that quantum number. As they are {\it a priori} 
indistinguishable, we label them by $\hat{L}_{\alpha} $ 
with the greek subscript being an (extended) flavor index going from $0$ to $3$.

The most general renormalizable superpotential for the supersymmetric
standard model (without R-parity) can be written then as
\begin{equation}
W = \varepsilon_{ab}\left[ \mu_{\alpha}  \hat{L}_{\alpha}^a \hat{H}_u^b
+ h_{ik}^u \hat{Q}_i^a   \hat{H}_{u}^b \hat{U}_k^{\scriptscriptstyle C}
+ \lambda_{i\alpha k}^{'} \hat{Q}_i^a \hat{L}_{\alpha}^b 
\hat{D}_k^{\scriptscriptstyle C} + \lambda_{\alpha \beta k}  \hat{L}_{\alpha}^a  
 \hat{L}_{\beta}^b \hat{E}_k^{\scriptscriptstyle C}
\right] + \lambda_{i jk}^{''}  \hat{D}_i^{\scriptscriptstyle C}  
\hat{D}_j^{\scriptscriptstyle C}  \hat{U}_k^{\scriptscriptstyle C}\; ,
\end{equation}
where  $(a,b)$ are $SU(2)$ indices, $(i,j,k)$ are the usual family (flavor) indices;
$\lambda$ and $\lambda^{''}$ are antisymmetric in the first two indices as
required by  $SU(2)$ and  $SU(3)$ product rules respectively,
though  only the former is shown explicitly here, 
$\varepsilon = \left({\begin{array}{cc} 0 & -1 \\ 1 &  0\end{array}}\right)$,
while the  $SU(3)$ indices are  suppressed.

At the limit where $\lambda_{ijk}, \lambda^{'}_{ijk},  \lambda^{''}_{ijk}$
and $\mu_{i}$  all vanish, one recovers the expression
for the R-parity preserving model, with $\hat{L}_{0}$ identified as $\hat{H}_d$. 
R-parity is exactly an {\it ad hoc} symmetry put in to make $\hat{H}_d$
stand out from the other $\hat{L}_i$'s. It is defined in terms of
baryon number, lepton number, and spin as, explicitly, 
${\mathcal R} = (-1)^{3B+L+2S}$. The consequence is that the accidental symmetries
of baryon number and lepton number in the standard model are preserved, at
the expense of making particles and superparticles having a categorically different
quantum number, R-parity. The latter is actually the most restrictive but not the
most effective discrete symmetry to control superparticle mediated proton 
decay\cite{IR}.

\section{ The single-VEV parametrization.}
With the above discussion, it is clear that in the phenomenology of
low energy supersymmetry, one approach worth studies is to take the complete
version of a supersymmetrized standard model without extra assumption and
check the phenomenological constraints on the various RPV
couplings. The large number of couplings make the task sound formidable.
For instance, the (color-singlet) charged fermion mass matrix is then given
by
\beq
{\cal{M}_{\scriptscriptstyle C}} =
 \left(
{\begin{array}{ccccc}
{M_{\scriptscriptstyle 2}} &  
\frac{g_{\scriptscriptstyle 2}{v}_{\scriptscriptstyle u}}{\sqrt 2} 
 & 0  & 0 & 0\\
 \frac{g_{\scriptscriptstyle 2}{v}_{\scriptscriptstyle d}}{\sqrt 2} &
 {{ \mu}_{\scriptscriptstyle 0}} &   
 -h^e_{i\scriptscriptstyle 1}  \frac{{v}_{\scriptscriptstyle i}}{\sqrt 2} 
 & -h^e_{i\scriptscriptstyle 2}  \frac{{v}_{\scriptscriptstyle i}}{\sqrt 2} 
 & -h^e_{i\scriptscriptstyle 3}  \frac{{v}_{\scriptscriptstyle i}}{\sqrt 2} 
 \\ 
 \frac{g_{\scriptscriptstyle 2}{v}_{\scriptscriptstyle 1}}{\sqrt 2} 
& {{ \mu}_{\scriptscriptstyle 1}} &
 h^e_{\scriptscriptstyle 1\!1}  \frac{{v}_{\scriptscriptstyle 0}}{\sqrt 2} 
+ 2\lambda_{\scriptscriptstyle 1i1}\frac{{v}_{\scriptscriptstyle i}}{\sqrt 2}&
 h^e_{\scriptscriptstyle 1\!2}  \frac{{v}_{\scriptscriptstyle 0}}{\sqrt 2} 
+ 2\lambda_{\scriptscriptstyle 1i2}\frac{{v}_{\scriptscriptstyle i}}{\sqrt 2}&
 h^e_{\scriptscriptstyle 1\!3}  \frac{{v}_{\scriptscriptstyle 0}}{\sqrt 2} 
+ 2\lambda_{\scriptscriptstyle 1i3}\frac{{v}_{\scriptscriptstyle i}}{\sqrt 2}\\
\frac{g_{\scriptscriptstyle 2}{v}_{\scriptscriptstyle 2}}{\sqrt 2} 
& {{ \mu}_{\scriptscriptstyle 2}} &
 h^e_{\scriptscriptstyle 2\!1}  \frac{{v}_{\scriptscriptstyle 0}}{\sqrt 2} 
+ 2\lambda_{\scriptscriptstyle 2i1}\frac{{v}_{\scriptscriptstyle i}}{\sqrt 2}&
 h^e_{\scriptscriptstyle 2\!2}  \frac{{v}_{\scriptscriptstyle 0}}{\sqrt 2} 
+ 2\lambda_{\scriptscriptstyle 2i2}\frac{{v}_{\scriptscriptstyle i}}{\sqrt 2}&
 h^e_{\scriptscriptstyle 2\!3}  \frac{{v}_{\scriptscriptstyle 0}}{\sqrt 2} 
+ 2\lambda_{\scriptscriptstyle 2i3}\frac{{v}_{\scriptscriptstyle i}}{\sqrt 2}\\
\frac{g_{\scriptscriptstyle 2}{v}_{\scriptscriptstyle 3}}{\sqrt 2} 
& {{ \mu}_{\scriptscriptstyle 3}} &
 h^e_{\scriptscriptstyle 3\!1}  \frac{{v}_{\scriptscriptstyle 0}}{\sqrt 2} 
+ 2\lambda_{\scriptscriptstyle 3i1}\frac{{v}_{\scriptscriptstyle i}}{\sqrt 2}&
 h^e_{\scriptscriptstyle 3\!2}  \frac{{v}_{\scriptscriptstyle 0}}{\sqrt 2} 
+ 2\lambda_{\scriptscriptstyle 3i2}\frac{{v}_{\scriptscriptstyle i}}{\sqrt 2}&
 h^e_{\scriptscriptstyle 3\!3}  \frac{{v}_{\scriptscriptstyle 0}}{\sqrt 2} 
+ 2\lambda_{\scriptscriptstyle 3i3}\frac{{v}_{\scriptscriptstyle i}}{\sqrt 2}
\end{array}}
 \right) \; ,
\eeq
from which the only definite experimental data are the three physical
lepton masses as the light eigenvalues, and the overall magnitude of
the electroweak symmetry breaking VEV. We must emphasize here that
the easier analysis of a model with only some small number of RPV
couplings admitted is, in general, lack of any theoretical motivation and
of very limited experimental relevance. Even the case of having only trilinear
RPV couplings in the superpotential is very difficult to motivate. Moreover, 
most studies of the type in the literature have extra assumptions about
the scalar potential or soft supersymmetry breaking terms which are usually
not explicitly addressed. This has led to quite some confusion and misleading
statements in literature of the subject.

It has been pointed out in Ref.\cite{I} that the single-VEV parametrization renders
the task of studying the complete theory of supersymmetry without R-parity
quite managable. The parametrization is nothing but an optimal choice of
flavor bases. In fact, doing phenomenological studies without specifying a choice 
of flavor bases is ambiguous. Recall that in quark physics of the standard
model, there are only 10 physical parameters from the 36 real parameters
of the two quark mass matrices written in a generic set of flavor bases. 
To standard model physics, the  26 extra parameters are absolutely
meaningless. Here for supersymmetry without R-parity, the choice of an optimal
parametrization mainly concerns the 4 $\hat{L}_\alpha$ flavors.
In the single-VEV parametrization,  flavor bases are chosen
such that: 1/ among $\hat{L}_\alpha$'s, only  $\hat{L}_0$, bears a VEV;
2/  $h^{e}_{ik} (\equiv 2\lambda_{i0k} =-2\lambda_{0ik}) 
=\frac{\sqrt{2}}{v_{\scriptscriptstyle d}}{\rm diag}
\{m_{\scriptscriptstyle 1},
m_{\scriptscriptstyle 2},m_{\scriptscriptstyle 3}\}$;
3/ $h^{d}_{ik} (\equiv \lambda^{'}_{i0k}) 
= \frac{\sqrt{2}}{v_{\scriptscriptstyle d}}{\rm diag}\{m_d,m_s,m_b\}$; 
4/ $h^{u}_{ik}=\frac{-\sqrt{2}}{v_u}V_{\scriptscriptstyle 
\!C\!K\!M}^{\dag} {\rm diag}\{m_u,m_c,m_t\}$. 
Under the parametrization, the (tree-level) mass 
matrices for {\it all} the fermions  {\it do not} involve any of
the trilinear RPV couplings though the
approach makes {\it no assumption} 
on any RPV coupling including even those from soft supersymmetry breaking;
and all the parameters used are uniquely defined. In fact, the above mass matrix
is reduced to the simple form :
\beq
{\mathcal{M}_{\scriptscriptstyle C}} =
 \left(
{\begin{array}{ccccc}
{M_{\scriptscriptstyle 2}} &  
\frac{g_{\scriptscriptstyle 2}{v}_{\scriptscriptstyle u}}{\sqrt 2}  
& 0 & 0 & 0 \\
 \frac{g_{\scriptscriptstyle 2}{v}_{\scriptscriptstyle d}}{\sqrt 2} & 
 {{ \mu}_{\scriptscriptstyle 0}} & 0 & 0 & 0 \\
0 &  {{ \mu}_{\scriptscriptstyle 1}} & {{m}_{\scriptscriptstyle 1}} & 0 & 0 \\
0 & {{ \mu}_{\scriptscriptstyle 2}} & 0 & {{m}_{\scriptscriptstyle 2}} & 0 \\
0 & {{ \mu}_{\scriptscriptstyle 3}} & 0 & 0 & {{m}_{\scriptscriptstyle 3}}
\end{array}}
\right)  \; .
\eeq
Each  $\mu_i$ parameter here characterizes directly the RPV effect
on the corresponding charged lepton  ($\ell_i = e$, $\mu$, and $\tau$).
For any set of other parameter inputs, the ${m}_i$'s can then be determined,
through a numerical procedure, to guarantee that the correct mass
eigenvalues of  $m_e$, $m_\mu$, and $m_\tau$  are obtained --- an issue
first addressed and solved in Ref.\cite{I}.

\section{Neutrino masses from the framework. }
Under the single-VEV parametrization, the tree-level neutral fermion 
(neutralino-neutrino) mass matrix  has  also
RPV contributions from the three $\mu_i$'s only. For the discussion below,
we write the mass matrix here as
\begin{equation} \label{mn}
{\cal{M}_{\scriptscriptstyle N}} 	
=  \left(
{\begin{array}{ccccccc}
{{M}_{\scriptscriptstyle 1}} & 0 &  \frac {{g}_{1}{v}_{u}}{2}
 &  -\frac{{g}_{1}{v}_{d}}{2} & 0 & 0 & 0 \\
0 & {{M}_{\scriptscriptstyle 2}} &  -\frac{{g}_{2}{v}_u}{2} & 
\frac{{g}_{2}{v}_{d}}{2} & 0 & 0 & 0 \\
 \frac {{g}_{1}{v}_{u}}{2} &   -\frac{{g}_{2}{v}_u}{2} & 0 & 
 - {{\mu}_{\scriptscriptstyle 0}} &  - {{ \mu}_{\scriptscriptstyle 1}} 
 &  - {{ \mu}_{\scriptscriptstyle 2}} &  - {{ \mu}_{\scriptscriptstyle 3}} \\
 -\frac{{g}_{1}{v}_{d}}{2} & \frac{g_{2}{v}_d}{2}
 &  - {{ \mu}_{0}} & W & 0 & Y & Z \\ 
0 & 0 &  - {{ \mu}_{\scriptscriptstyle 1}} & 0 & 0 & 0 & 0 \\
0 & 0 &  - {{ \mu}_{\scriptscriptstyle 2}} & Y & 0 & A & C \\
0 & 0 &  - {{ \mu}_{\scriptscriptstyle 3}} & Z & 0 & C & B
\end{array}}
 \right)  \; ,
\end{equation}
with parameters $A,\; B,$
and $C$, and $W,\; Y,$ and $Z$ being two groups of relevant 1-loop
contributions to be addressed. Setting all these to zero retrieves
the tree-level result where an admixture of the three
neutral fermionic states from the $\hat{L}_i$'s gets a nonzero mass
from mixing with the gauginos and higgsinos. Note that at the limit of small 
$\mu_i$'s, the neutral states
correspond to $\nu_e$, $\nu_\mu$, and $\nu_\tau$.

An important question is whether the $\mu_i$'s are large or small. A careful
analysis of an exhaustive list of constraints from tree-level leptonic
phenomenology illustrates that while $\mu_{\scriptscriptstyle 1}$ has to be
small, $\mu_{\scriptscriptstyle 2}$ and, especially, 
$\mu_{\scriptscriptstyle 3}$ do not have to\cite{II}. In fact, MeV scale neutrino 
mass is easily admitted, with interesting implications on lepton number 
violating processes. Fitting neutrino oscillation data will then call
for extensions of the model. We are interested here in the complementary
scenario of sub-eV neutrino mass(es). In that case, the 1-loop contributions
could also be significant. Explicitly, we assume a three neutrino scenario
motivated by the recent zenith angle dependence measurement by the 
Super-K experiment\cite{SuperK}. There have been a lot
of studies on the topic, details on which we are not going into here\cite{nu}.
The scenario is summarized by
\begin{eqnarray*}
\Delta\!m^2_{\rm atm} &\simeq& (0.5-6)\times 10^{-3}\, \mbox{eV}^2 \\
\sin^2\!2\theta_{\rm atm}   &\simeq& (0.82-1) \\
  \Delta\!m^2_{\rm sol} &\simeq& (4-10)\times 10^{-6}\, \mbox{eV}^2 \\
\sin^2\!2\theta_{\rm sol}   &\simeq& (0.12-1.2)\times 10^{-2}
\end{eqnarray*}
with $\nu_{\scriptscriptstyle \mu} - \nu_{\scriptscriptstyle \tau}$ to be 
responsible for the Super-K atmospheric 
result and MSW-oscillation of $\nu_e$ for the solar neutrino problem. 
The most natural setting then would be for the two neutrino mass
eigenvalues of the $\nu_\mu - \nu_\tau$ system to have $m^2 \simeq
  \Delta\!m^2_{\rm sol}$ and $  \Delta\!m^2_{\rm atm}$. We will
concentrate on this particular scenario below. Our concern will be focused on 
the compatibility of the required maximal mixing between 
$\nu_{\scriptscriptstyle \mu}$ and $\nu_{\scriptscriptstyle \tau}$, with the 
general hierarchical flavor structure of the quarks and charged leptons.

Consider
${\cal{M}_{\scriptscriptstyle N}}$ of Eq.(\ref{mn}) in the $3+4$ block form
$\left(
\begin{array}{cc}
\cal{M} & \xi^{\scriptscriptstyle T} \\
\xi & m_\nu^{\scriptscriptstyle 0}
\end{array}
 \right)$. 
For small $\mu_i$'s, it has a ``see-saw"  type structure,  with the effective
neutrino mass matrix given by
\begin{equation} \label{ss}
m_\nu = - \xi {\cal{M}}^{-1}  \xi^{\scriptscriptstyle T} + 
m_\nu^{\scriptscriptstyle 0} \; .
\end{equation}
In the case that the $\mu_i$ contributions dominate, the first term of the
equation gives
\beq
m_\nu =
-  \frac {1}{2}
\frac{  {v}^{2} \cos^2\!\!\zb 
\left( x{g}_{\scriptscriptstyle 2}^{2} 
+ {g}_{\scriptscriptstyle 1}^{2} \right) }
{\mu_{\scriptscriptstyle 0} \left[ 2xM_{\scriptscriptstyle 2}
 \mu_{\scriptscriptstyle 0} -
 \left( x{g}_{\scriptscriptstyle 2}^{2}+{g}_{1\scriptscriptstyle }^{2}\right) 
{v}^2 \sin\!{\zb}\cos\!{\zb} \right] } 
\left(
\begin{array}{cc}
\mu_{\scriptscriptstyle 2}^2
& \mu_{\scriptscriptstyle 2} \mu_{\scriptscriptstyle 3} \\
\mu_{\scriptscriptstyle 2} \mu_{\scriptscriptstyle 3}
 & \mu_{\scriptscriptstyle 3}^2
\end{array}
 \right) \; ,
\eeq 
where we have neglected contributions involving the $\nu_{\scriptscriptstyle e}$ 
state and hence shrinked the matrix to $2\times 2$. Dropping the pre-factor,
the matrix is diagonalized by a rotation of
$\tan\theta= \mu_{\scriptscriptstyle 2} / \mu_{\scriptscriptstyle 3}$,
giving eigenvalues $0$ and 
$\mu_{\scriptscriptstyle 2}^2+ \mu_{\scriptscriptstyle 3}^2$. For this to fit in
our neutrino oscillation scenario, it requires
$\sqrt{\mu_{\scriptscriptstyle 2}^2+ \mu_{\scriptscriptstyle 3}^2}
\cos\!{\beta} \sim 10^{-4}\, \mbox{GeV}$
and ${{\mu_{\scriptscriptstyle 2}}}/ {{\mu_{\scriptscriptstyle 3}}} 
\mathrel{\raise.3ex\hbox{$>$\kern-.75em\lower1ex\hbox{$\sim$}}}  0.6358 $.
It is interesting to note that the structure of matrix in the form 
$\left(
\begin{array}{cc}
a^2 & ab \\ ab & b^2
\end{array}
 \right)$ 
naturally admits maximal mixing with a hierarchy in mass eigenvalues.

There are two types of 1-loop contributions to Eq.(\ref{mn}) or 
$m_\nu^{\scriptscriptstyle 0} $ of Eq.(\ref{ss}) --- the quark-squark
 and  lepton-slepton loops. The former is given by
\beq \label{LL}
(m^{\scriptscriptstyle LL})^{q}_{\alpha \beta} = \frac{3}{16\pi^2} 
\lambda^{'}_{i\alpha j} \lambda^{'}_{j \beta i} \left(
\frac{A^d_{j}m^d_{i}}{\tilde{m}^2_{q_j}} + 
 \frac{A^d_{i}m^d_{j}}{\tilde{m}^2_{q_i}} \right) \; ,
\eeq
where $m^{\scriptscriptstyle LL}$ corresponds to the lower $4\times 4$
block of  Eq.(\ref{mn}), and the soft supersymmetry breaking
trilinear terms $A^d$ are assumed to be dominately diagonal, as generally
expected. We get the dominating contribution to 
$m_\nu^{\scriptscriptstyle 0}$ as
\beq \label{0q}
(m_\nu^{\scriptscriptstyle 0})^q \simeq \frac{3}{8\pi^2}
\frac{m^2_b}{M_{\scriptscriptstyle S\!U\!S\!Y}}
\left( \begin{array}{cc}
 \lambda^{'2}_{\scriptscriptstyle 3\!2\!3} & 
\lambda^{'}_{\scriptscriptstyle 3\!2\!3}
 \lambda^{'}_{\scriptscriptstyle 3\!3\!3} \\
\lambda^{'}_{\scriptscriptstyle 3\!2\!3} 
\lambda^{'}_{\scriptscriptstyle 3\!3\!3} & 
 \lambda^{'2}_{\scriptscriptstyle 3\!3\!3}
\end{array}\right) \; .
\eeq
If this contribution dominates, we have a mass matrix with the same general
structure as the $\mu_i$ dominating case above, hence again  natural
maximal mixing with a hierarchy in mass eigenvalues. It requires
$\lambda^{'} \sim 10^{-4}$ and 
${\lambda^{'}_{\scriptscriptstyle 3\!2\!3}}/
{ \lambda^{'}_{\scriptscriptstyle 3\!3\!3}} \mathrel{\raise.3ex\hbox{$>$\kern-.75em\lower1ex\hbox{$\sim$}}}  0.6358$. However, it is important to note that
the natural structure would be spoiled if the $\mu_i$ contribution and
the present one are at about the same level. Finally, we note also that the
$4\times 4$ form of Eq.(\ref{LL}) allows one to check that the contributions to
the $W$, $Y$, and $Z$ entries of Eq.(\ref{mn}) are really negligible.

The lepton-slepton loop contributions have a different structure. We have,
similar to the the previous case,
\beq  
(m^{\scriptscriptstyle LL})^\ell_{\alpha \beta} = \frac{1}{16\pi^2} 
\lambda_{i \alpha j} \lambda_{j \beta i} \left(
\frac{A^{\ell}_{j}m^\ell_{i}}{\tilde{m}^2_{\ell_j}} + 
\frac{A^{\ell}_{i}m^\ell_{j}}{\tilde{m}^2_{\ell_i}} \right)  \; .
\eeq
In this case, however, the antisymmetry in 
$\lambda \hat{L}\hat{L}\hat{E}^{\scriptscriptstyle c}$ couplings between 
the two $\hat{L}$'s gives the dominating contribution as
\beq \label{0l}
(m_\nu^{\scriptscriptstyle 0})^\ell \!\simeq \!\frac{1}{8\pi^2
M_{\scriptscriptstyle S\!U\!S\!Y}}\!
\left(\!\!\! \begin{array}{cc}
\!m_\tau^2 \lambda_{\scriptscriptstyle 3\!2\!3}^2 \!\!&
\!\!- m_\mu m_\tau \lambda_{\scriptscriptstyle 3\!2\!2}
 \lambda_{\scriptscriptstyle 3\!2\!3}  \\
\!- m_\mu m_\tau \lambda_{\scriptscriptstyle 3\!2\!2}
 \lambda_{\scriptscriptstyle 3\!2\!3}
\!\!& \!\! m_\mu^2 \lambda_{\scriptscriptstyle 3\!2\!2}^2
\!\end{array}
\!\!\!\right)\; ,
\eeq
which is in general incompatible with large mixing. To fit in the neutrino 
oscillation scenario, we would hence like the lepton-slepton loops to play
a secondary role, which requires $\lambda$'s of order $10^{-4}$ or less.

\section{Flavor structure among the 4 $\hat{L}_\alpha$'s}
After the above discussion of the various sources of neutrino masses, 
let us look at the flavor stucture more carefully. We will adopt a flavor
model independent approach along the idea of the approximate 
flavor symmetry\cite{H}. The idea is to 
attach a suppression factor to each chiral multiplet.
For example, the down-quark mass matrix would looks like
\[
{\cal M}_d = \left( \begin{array}{ccc}
\varepsilon_{\!\scriptscriptstyle Q_1}
\varepsilon_{\!\scriptscriptstyle D^c_1} &
\varepsilon_{\!\scriptscriptstyle Q_1}
\varepsilon_{\!\scriptscriptstyle D^c_2} &
\varepsilon_{\!\scriptscriptstyle Q_1}
\varepsilon_{\!\scriptscriptstyle D^c_3} \\
\varepsilon_{\!\scriptscriptstyle Q_2}
\varepsilon_{\!\scriptscriptstyle D^c_1} &
\varepsilon_{\!\scriptscriptstyle Q_2}
\varepsilon_{\!\scriptscriptstyle D^c_2} &
\varepsilon_{\!\scriptscriptstyle Q_2}
\varepsilon_{\!\scriptscriptstyle D^c_3} \\
\varepsilon_{\!\scriptscriptstyle Q_3}
\varepsilon_{\!\scriptscriptstyle D^c_1} &
\varepsilon_{\!\scriptscriptstyle Q_3}
\varepsilon_{\!\scriptscriptstyle D^c_2} &
\varepsilon_{\!\scriptscriptstyle Q_3}
\varepsilon_{\!\scriptscriptstyle D^c_3}
\end{array} \right)
\]
with the flavor hierarchy
$ \varepsilon_{\!\scriptscriptstyle Q_1} \ll
\varepsilon_{\!\scriptscriptstyle Q_2} \ll
\varepsilon_{\!\scriptscriptstyle Q_3}$ 
and
$\varepsilon_{\!\scriptscriptstyle D^c_1} \ll
\varepsilon_{\!\scriptscriptstyle D^c_2} \ll
\varepsilon_{\!\scriptscriptstyle D^c_3}$.
Diagonalization gives
$m_{\scriptscriptstyle d} :
m_{\scriptscriptstyle s} :
m_{\scriptscriptstyle b} =\varepsilon_{\!\scriptscriptstyle Q_1}
\varepsilon_{\!\scriptscriptstyle D^c_1} :
\varepsilon_{\!\scriptscriptstyle Q_2}
\varepsilon_{\!\scriptscriptstyle D^c_2} :
\varepsilon_{\!\scriptscriptstyle Q_3}
\varepsilon_{\!\scriptscriptstyle D^c_3}$
with mixings given by factors of the form
$ {\varepsilon_{\!\scriptscriptstyle Q_i} }/
{\varepsilon_{\!\scriptscriptstyle Q_j}}$.
We adopt the approach here for two major reasons. First of all, while an explicit
flavor model may be designed to contain very specific features needed to
reconcile with experimental number, the approach emphasizes on generic
flavor structure features which would fit in easily any natural flavor model.
If the approach can easily accomodate the required ``smallness" of various 
RPV couplings, it builds a strong case for the latter couplings to be 
considered on the same footing as the R-parity conserving ones. Second, it is
clear, from the above discussions, that we are dealing with a large number
of parameters but a small amount of data. In such a situation, detailed model
construction is very unlikely to be fruitful. We would rather take a humble
approach and discuss issues that will not be easily washed away when more
data becomes available.

In the small $\mu_i$ case considered, the $\hat{L}_i$ basis under 
the single-VEV parametrization gives excellent alignment with the
charged lepton mass eigenstate basis. However, going into the approximate
flavor symmetry perspective, we have to start with generic, non-diagonal, flavor 
bases. The mis-alignment between the two is a major problem hindering a
complete discussion of the flavor structure here. This is tired up with the 
question of the natural values of our $\mu_i$'s. Careful analysis of the 
scalar potential and vacuum solution is needed to settle the issue. We 
will leave that to a future studies while trying to learn something from
a naive analysis.

It is easy to see that our neutrino oscillation scenario, together with the 
known charged lepton masses, suggests
\begin{eqnarray*}
\varepsilon_{\!\scriptscriptstyle L_1} \ll
\varepsilon_{\!\scriptscriptstyle L_2} \sim
\varepsilon_{\!\scriptscriptstyle L_3} \ll
\varepsilon_{\!\scriptscriptstyle L_0} \; , \\
\noalign{\hbox{and}}
\varepsilon_{\!\scriptscriptstyle E^c_1} \ll
\varepsilon_{\!\scriptscriptstyle E^c_2} \ll
\varepsilon_{\!\scriptscriptstyle E^c_3}  \; .
\end{eqnarray*}
With 
$\varepsilon_{\!\scriptscriptstyle L_2} \sim
\varepsilon_{\!\scriptscriptstyle L_3}$, however, the factors that go 
with $L_2$ and $L_3$ (after a diagonalizing rotaton is taken into consideration)
would be $\sin\!2\theta_{23} 
\sqrt{\varepsilon_{\!\scriptscriptstyle L_2}^2 + 
\varepsilon_{\!\scriptscriptstyle L_3}^2}$ and $\cos\!2\theta_{23} 
\sqrt{\varepsilon_{\!\scriptscriptstyle L_2}^2 + 
\varepsilon_{\!\scriptscriptstyle L_3}^2}$. These are, of course,
still of the same order of magnitude as 
$\varepsilon_{\!\scriptscriptstyle L_2}$ and 
$\varepsilon_{\!\scriptscriptstyle L_3}$.
If we take $\varepsilon_{\!\scriptscriptstyle E^c_3}\sim 1$, as we do with
other third family factors such as $\varepsilon_{\!\scriptscriptstyle Q_3}$,
we would use
\[
\cos\!2\theta_{\scriptscriptstyle 23} 
\sqrt{\varepsilon_{\!\scriptscriptstyle L_2}^2 + 
\varepsilon_{\!\scriptscriptstyle L_3}^2} \sim \frac{m_\tau}{m_t}
\]
to fix the $\tau$ mass. That is the maximal suppression in the 
$L_3$ flavor factor we can have. If we further take
\[
\varepsilon_{\!\scriptscriptstyle D^c_3} \sim \frac{m_b}{m_t} \; ,
\]
we would have naturally, at  
$M_{\scriptscriptstyle S\!U\!S\!Y} = 100\, \mbox{GeV}$
[ {\it cf. } Eqs.(\ref{0q}) and (\ref{0l})],
	$\lambda^{'}_{\scriptscriptstyle 3\!3\!3} \sim 
\lambda^{'}_{\scriptscriptstyle 3\!2\!3} \sim 5\times 10^{-4} $,
$\lambda_{\scriptscriptstyle 3\!2\!3} \sim 10^{-4} $,
and $\lambda_{\scriptscriptstyle 3\!2\!2} \sim 10^{-5} $.
Hence, amazing enough, a bit larger value of $M_{\scriptscriptstyle S\!U\!S\!Y}$
(squark and slepton masses) would gives the quark-squark loop dominating scenario naturally.

To fit our neutrino oscillation scenario with  $\mu_i$'s being
the dorminating neutrino mass contribution will require a higher
$M_{\scriptscriptstyle S\!U\!S\!Y}$ 
and $\mu_{\scriptscriptstyle 1} \ll \mu_{\scriptscriptstyle 2}
\sim \mu_{\scriptscriptstyle 3} \ll \mu_{\scriptscriptstyle 0}$
with $\mu_{\scriptscriptstyle 3}/ \mu_{\scriptscriptstyle 0} < 10^{-6}$.
Feasibility of this case we cannot judge, as mentioned above, until the 
complicated analysis of the scalar potential has been performed.

\section{Summary.}
In summary, from our brief analysis here, we have illustrated a few
interesting issues in the flavor structure of supersymmetry without R-parity.
The question of the natural suppression of the $\mu_i$'s is important.
It is however a subtle issue which has to be analyzed from a careful
study of  the full five-doublet ($4\;\hat{L}_\za + \hat{H}_u$) scalar 
potential with the most generic soft supersymmetry breaking terms. 
Assuming that can be explained, we illustrate above that 
the suppressed values of the RPV couplings, required for fitting the
limiting scenario of neutrino oscillations motivated by the recent
Super-K result,  fit very well  into
an approximate flavor symmetry perspective. Success of the latter
is a strong indication that the R-parity (or lepton number) violating
couplings  are ``naturally" small, as the light fermion masses are,
and their explanation most probably lies under a common theory of
flavor structure.

{\bf Acknowledgement :}
A. Mitov is to be thanked for reading the manuscript.
This work was supported in part by the U.S. Department of Energy,
under grant DE-FG02-91ER40685

\end{document}